\documentclass[10pt]{iopart}

\usepackage{graphicx}
\usepackage{amssymb}
\usepackage{dcolumn}
\usepackage{bm}
\usepackage{color}

\begin{document}

\title{Evolution of the Fermi surface of BiTeCl with pressure}

\author{D.\ VanGennep$^1$, D.\ E.\ Jackson$^1$, D.\ Graf$^2$, H.\ Berger$^3$, and J.\ J.\ Hamlin$^1$}

\address{$^1$Department of Physics, University of Florida, Gainesville, FL 32611\\
	  $^2$National High Magnetic Field Laboratory, Tallahassee, FL 32310\\
	  $^3$Institute of Condensed Matter Physics, \'{E}cole Polytechnique F\'{e}d\'{e}rale de Lausanne, CH-1015 Lausanne, Switzerland}

\ead{jhamlin@ufl.edu}

\begin{abstract}
We report measurements of Shubnikov-de Haas oscillations in the giant Rashba semiconductor BiTeCl under applied pressures up to $\sim 2.5\,\mathrm{GPa}$. We observe two distinct oscillation frequencies, corresponding to the Rashba-split inner and outer Fermi surfaces. BiTeCl has a conduction band bottom that is split into two sub-bands due to the strong Rashba coupling, resulting in two spin-polarized conduction bands as well as a Dirac point. Our results suggest that the chemical potential lies above this Dirac point, giving rise to two Fermi surfaces. We use a simple two-band model to understand the pressure dependence of our sample parameters. Comparing our results on BiTeCl to previous results on BiTeI, we observe similar trends in both the chemical potential and the Rashba splitting with pressure.  
\end{abstract}

\maketitle
\ioptwocol

\section{Introduction}
In materials with a large spin orbit interaction, the Rashba effect can lead to electronic states that are spin-split in zero magnetic field~\cite{Rashba_1960,Bychkov_1984_1}.  The strength of the splitting is described by the Rashba parameter, $\alpha$, where the dispersion of the two spin-polarized states is described by $E^{\pm}(\textbf{k}) = (\hbar ^2 \textbf{k}^2/2m^*) \pm \alpha |\textbf{k}|$.  The effect can derive either from perpendicular electric fields at an interface or from inversion asymmetry.  The discovery of the Rashba effect was followed by a proposal to utilize the effect to build a spin transistor~\cite{Datta1990}, contributing substantially to the development of the field of spintronics~\cite{Bihlmayer_2015_1}.

Many efforts to apply the Rashba effect have focused on surfaces and interfaces because the strength of the splitting ($\alpha$) due to inversion asymmetry is too small to be useful in most semiconducting materials.  This changed with the discovery of a so-called ``giant'' Rashba effect in the bulk of the semiconductor BiTeI~\cite{Ishizaka2011}.  In BiTeI, $\alpha \sim 4$\,eV\AA, substantially larger than had previously been found in systems exhibiting the Rashba effect.  The closely related compounds BiTeBr and BiTeCl have similarly large values of the Rashba parameter with $\alpha \sim 2$\,eV\AA~\cite{Moreschini2015}.

BiTe$X$ ($X$ = Cl, Br, I) compounds were first synthesized and structurally characterized by Shevelkov \textit{et al}.~\cite{Shevelkov_1995_1}.  All three compounds crystallize in slightly different variants of a non-centrosymmetric layered hexagonal structure.  All three materials appear to be small band-gap semiconductors, where the strong spin-orbit interaction deriving from the presence of heavy ions leads to a tendency towards band inversion.  In the case of BiTeI, the band inversion only occurs at elevated pressures~\cite{bahramy_2011_2}, while for BiTeCl the band inversion is present at ambient pressure~\cite{chen_2013_2}.  The band inversion is expected to produce an unusual topologically non-trivial state in these materials.  Angle resolved photoemission spectroscopy (ARPES) measurements have confirmed the presence of topological surface states in BiTeCl at ambient pressure~\cite{chen_2013_2}.  Due to the inversion asymmetry of the crystal structure, BiTeCl appears to belong to a unique class of topological insulators in which the surface state charge carriers are $n$-type on one side of the material and $p$-type on the other.

Although nominally insulating, non-stoichiometry or site disorder leads these materials to have a well defined Fermi surface.  Several reports have detailed measurements of quantum oscillations in the conductivity (Shubnikov-de Haas oscillations) for BiTe$X$ compounds.  When the magnetic field is applied parallel to the crystallographic $c$-axis, oscillations are visible at either one or two frequencies, depending on how close the chemical potential is to the Dirac point in the bulk electronic structure that derives from the Rashba splitting.  Note that this Dirac point is distinct from the one associated with the topological surface states.  Chen \textit{et al}.~\cite{Chen_2014_1} performed a quantum oscillation study on BiTeCl single crystals grown using a self-flux method and observed one 3D Fermi surface with $n$-type carriers. Further quantum oscillation and infrared studies were performed on single crystals of BiTeCl which had a chemical potential above the bulk Dirac point, giving rise to two sets of quantum oscillations originating from one small, inner Fermi surface (IFS) and a larger, outer Fermi surface (OFS)~\cite{Martin_2014_1}.  Both sets of oscillations (IFS and OFS) exhibit a nontrivial Berry's phase of $\pi$~\cite{murakawa_2013_1,Xiang_2015_1}. 

By tuning the chemical potential in BiTe$X$ compounds, it is possible to drive an electronic topological transition (ETT) from one Fermi surface to two Fermi surfaces, which occurs when the Fermi level crosses the Dirac point.  When the chemical potential lies below the Dirac point, the Fermi surface takes on a torus shape, while for a chemical potential above the Dirac point, two Fermi surfaces exist~\cite{Xiang_2015_1}.  A quantum oscillation study used multiple BiTeCl samples with varying carrier density to show that the ETT can be driven by chemical means~\cite{Xiang_2015_1}.  In the case of BiTeI, it was shown that the ETT can be driven by pressure~\cite{vangennep_2014_1}.

The properties of BiTe$X$ compounds at elevated pressures have been reported in a number of papers.  Such work was initially motivated by the prediction of a pressure-driven quantum phase transition from topologically trivial to non-trivial in BiTeI at 2-4 GPa~\cite{bahramy_2011_2}.  A room temperature infrared and x-ray diffraction study found that BiTeI remains in the ambient pressure structure to 8 GPa and found evidence for the band inversion occurring in the pressure range 2-3 GPa~\cite{xi_2013_1}.  Two Shubnikov-de Haas studies found evidence that, at low temperatures, the band inversion had yet to occur at pressures below 2 GPa~\cite{vangennep_2014_1} and 3 GPa~\cite{Ideue2014}.  A subsequent quantum oscillation study found that deviations in the phase of the quantum oscillations with pressure were consistent with band structure calculations that predicted a change in the curvature of the Fermi surface at the critical pressure for the band inversion~\cite{Park2015}.  Together, these studies suggest that bulk signatures of the band inversion may be rather subtle in these materials.  At substantially higher pressures, BiTe$X$ compounds undergo transitions to different crystal structures.  The Cl, Br, and I, variants have all been found to exhibit superconductivity in one or more of their high pressure structures~\cite{Ying_2016_1,Qi2017,VanGennep2017}. 

Electrical transport measurements are well suited to moderately high pressure conditions and can provide a wealth of information on the influence of pressure~\cite{errandonea_2014,hamlin_2015,shimizu_2015}.  To investigate the effects of external pressure on BiTeCl, we performed both Hall and Shubnikov-de Haas quantum oscillation measurements on single crystals of BiTeCl under hydrostatic pressures up to $\sim 2.5\,\mathrm{GPa}$. At these pressures, BiTeCl remains in the ambient pressure crystal structure.  With the use of a simple, two-band model~\cite{vangennep_2014_1}, we use the measured band masses and oscillation frequencies to calculate band parameters related to the Rashba splitting. Comparing the results of this study to previous results on BiTeI~\cite{vangennep_2014_1}, we find similar trends in both the chemical potential and the Rashba splitting with pressure.

\section{Experimental methods}
 Single crystals of BiTeCl were grown by the chemical vapor transport method~\cite{jacimovic_2014_1}.
 Small pieces of sample with dimensions of about $500\,\mathrm{\mu m} \times 500\,\mathrm{\mu m} \times 50\,\mathrm{\mu m}$ were cut from a larger crystal. Pt wires were connected to the samples using EPO-TEK H20E conductive epoxy. The samples were then mounted to the wire and fiber optic feed-throughs of two piston-cylinder type pressure cells. One cell was constructed from MP35N alloy, and the other was a hybrid cell consisting of an inner cylinder of MP35N and an outer cylinder of BeCu. The pressure was calibrated at both room temperature and the lowest temperature reached, using the fluorescence of the R1 peak of a small ruby chip~\cite{piermarini_1975_1,syassen_2008}. Note that up to a few GPa, the early and more modern ruby pressure calibrations are in good agreement.  Daphne 7474 oil was used as the pressure-transmitting medium surrounding the sample~\cite{murata_2008_1}. Daphne 7474 remains liquid at room temperature for pressures below 3.7 GPa, ensuring nearly hydrostatic conditions in our experiments, since pressure was always applied or released at room temperature.
 
 Resistance and Hall measurements were performed from $\sim 2-300\,\mathrm{K}$ in either a $16\,\mathrm{T}$ Quantum Design PPMS or an $18\,\mathrm{T}$ Oxford magnet with a variable temperature insert. Four-wire resistance measurements were performed in the crystalline $ab$-plane using either a Quantum Design PPMS resistance bridge or a Lakeshore 370 resistance bridge.  Magnetic fields were applied along the $c$-axis.  All of the samples in this study were cut from the same parent crystal. Samples 3, 4, 5, and 7 were studied at ambient pressure only, while samples 2, 8, and 9 were chosen for  measurements under pressure. All samples were wired in the van der Pauw geometry, allowing for Hall measurements, with the exception of sample 2, which was wired in the standard bar configuration. 
 
\section{Results}
\begin{figure}
    \includegraphics[width=\columnwidth]{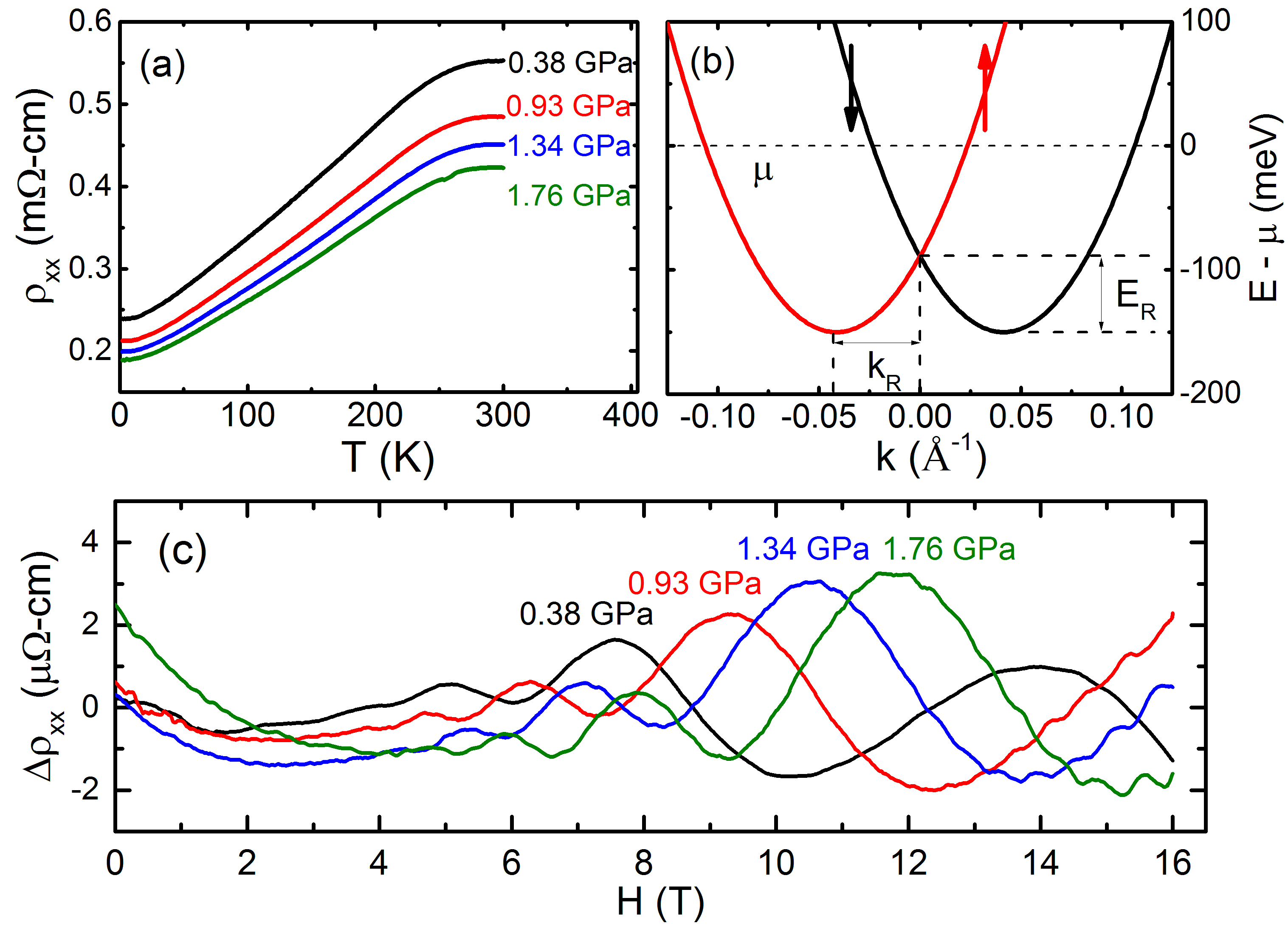}
    \caption{(a) Electrical resistivity of sample 8 in the $ab$-plane versus temperature, measured from 2-300 K at several pressures. (b) Conduction bands of sample 8 at 0.93 GPa, extracted from oscillation data using the analysis described in Ref.~\cite{vangennep_2014_1}. Red and black arrows denote the spin-polarization of each band. (c) The oscillatory part of the magnetoresistance of sample 8 at various pressures.}
 \label{fig1}
\end{figure}

All of the samples examined yielded very similar results. An overview of our results from sample 8 is presented in Fig.~\ref{fig1}. An increase in pressure results in a decrease in the resistivity, but the overall trend in the temperature dependence of the resistivity does not change. Relatively large, low frequency quantum oscillations are visible down to $\sim 4\,\mathrm{T}$ and up to the maximum field. High frequency quantum oscillations stemming from the outer Fermi surface are only visible down to $\sim 10\,\mathrm{T}$. Hall measurements reveal the samples are $n$-type, with total carrier densities for samples 8 and 9 that are both on the order of $10^{19}~ \mathrm{cm^{-3}}$ and increase monotonically with pressure by $\sim 20\, \mathrm{\%}$ from $\sim 0.5 - 2.0\,\mathrm{GPa}$. The large carrier densities found in our samples indicate that the chemical potential lies above the Dirac point~\cite{Xiang_2015_1}. No significant third-order term is present in the data for $\rho_{xy}$ vs $H$.

The quantum oscillations were analyzed by first subtracting a polynomial background from the magnetoresistance. The frequencies were then extracted from the resulting data using a fast Fourier transform (FFT). The frequencies have also been determined by taking the slope of a plot of the Landau level index, $n$, versus the value of $1/B$ where the oscillatory part of the magnetoresistance passes through a maximum. Both methods of extracting the frequency yield nearly identical results. For sample 8 and 9, both $\rho_{xx}$ and $\rho_{xy}$ were measured, and $\rho_{xx} > \rho_{xy}$ for our entire range of magnetic field at low temperatures, allowing for the assignment of Landau index, $n$, integers to the maxima in the longitudinal magnetoresistance. 

\begin{figure}
  \raggedleft
    \includegraphics[width=\columnwidth]{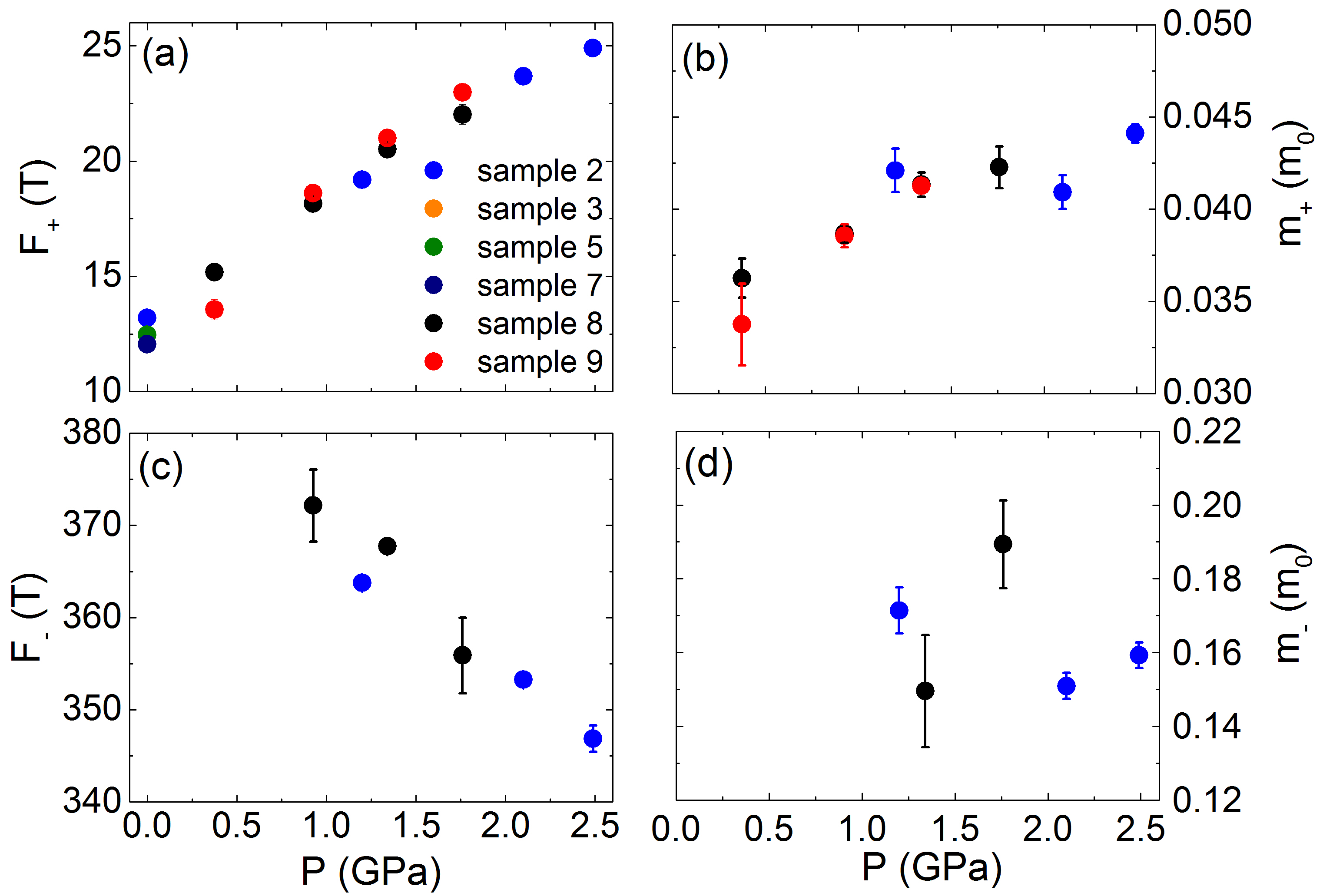}
    \caption{Fermi surface parameters at 2 K as a function of pressure. (a) The IFS frequency, $F_+$, rises monotonically with pressure, (b) the IFS mass, $m_+$, trends upward, (c) the OFS frequency, $F_-$, decreases, while (d) the OFS mass, $m_-$ remains roughly constant.}
 \label{fig2}
\end{figure}

The temperature dependence of the amplitude of the oscillations was used to extract the effective masses of both the OFS and IFS ~\cite{Shoenberg1984}. Figure~\ref{fig2} presents the pressure dependence of the IFS and OFS cyclotron frequencies and masses for all of the samples studied. The OFS oscillations are only resolvable for samples 2 and 8. For all samples, the IFS frequency, $F_{+}$, increases monotonically with pressure, while the OFS frequency, $F_{-}$, decreases. The IFS cyclotron mass, $m_{+}$, tends to increase with pressure for all of our samples, while the OFS cyclotron mass, $m_{-}$, stays roughly constant.

\begin{figure}
  \raggedleft
    \includegraphics[width=\columnwidth]{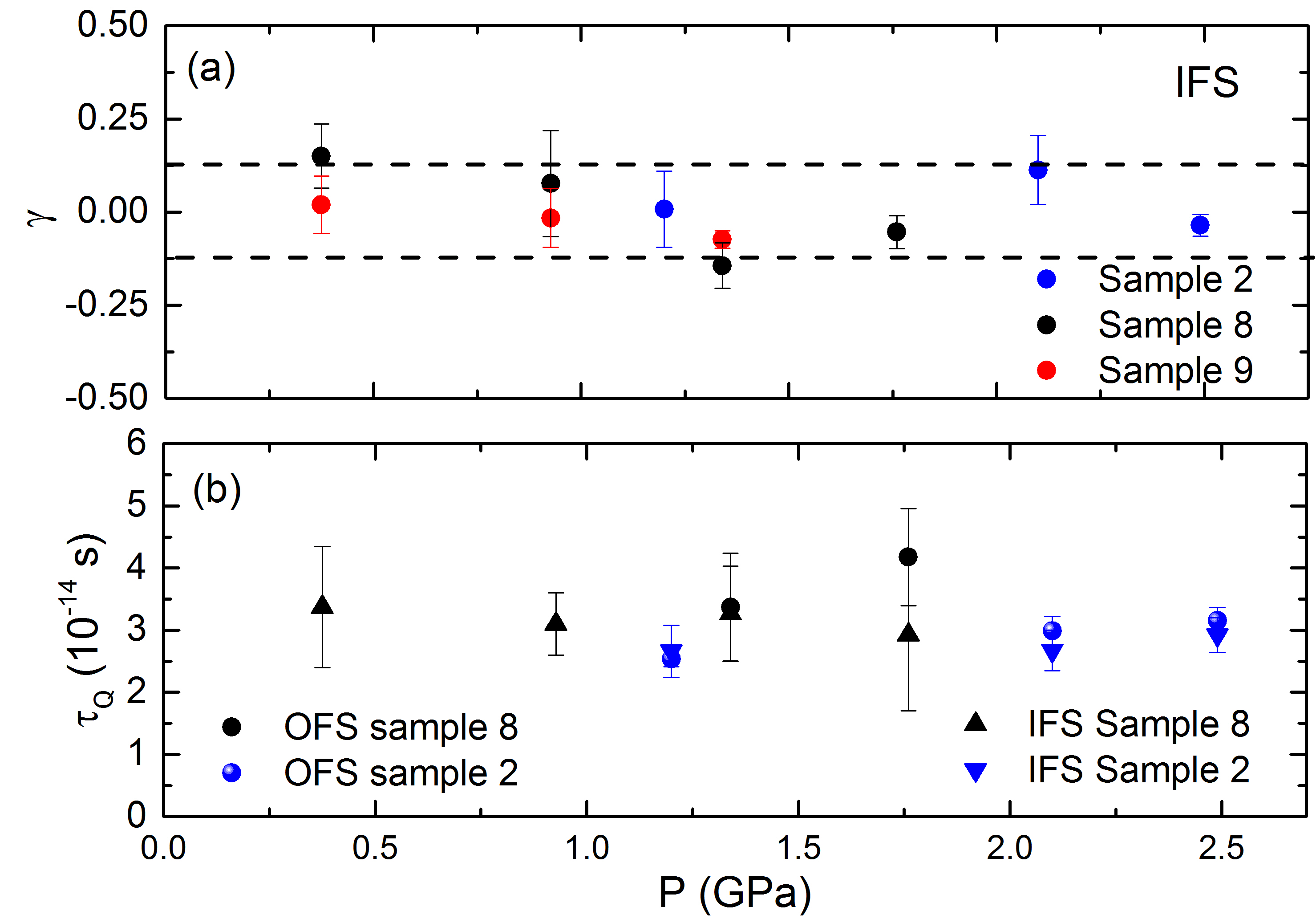}
    \caption{(a) Quantum oscillation phase, $\gamma$, of the IFS as a function of pressure. Horizontal dashed lines correspond to $\phi_B = \pi$ and $\delta = \pm1/8$.  (b) Quantum scattering time, $\tau _Q$, of both the IFS and OFS as a function of pressure.}
 \label{fig3}
\end{figure}

Figure~\ref{fig3} shows our results for the phase of the quantum oscillations, $\gamma$, associated with the IFS as well as the quantum scattering times, $\tau_Q$, for both the IFS and OFS. The phase was obtained from the $n$-intercept of the Landau level plots. Since the maximum fields in these experiments are below the quantum limit for the IFS, we have shifted the values of $n$ by the integer value that produces the smallest intercept. The value of $\phi_B$ can be calculated from $\gamma$ using the relation: $\gamma = 1/2 - \phi_B/{2\pi} + \delta$, where $\delta = \pm1/8$ for a 3D Fermi surface \cite{Shoenberg1984, Mikitik1999}. For all pressures, the IFS exhibits a non-trivial Berry's phase of $\pi$, resulting from the helical spin texture produced by the Rashba spin-splitting of the bulk~\cite{murakawa_2013_1}.  The OFS Landau level plots require too large of an extrapolation to extract a reliable $\gamma$. The field dependence of the oscillation amplitudes were used to extract the Dingle temperatures, $T_D$, and thus, the quantum scattering time $\tau_Q$ using $T_D = \hbar/{2k_B\tau_Q}$. We observe the IFS $\tau_Q$ remaining roughly constant throughout our range of pressures, while the OFS $\tau_Q$ may slightly increase for both samples 2 and 8.

\section{Discussion}
\label{sec:Discussion}
\begin{figure}
  \raggedleft
    \includegraphics[width=\columnwidth]{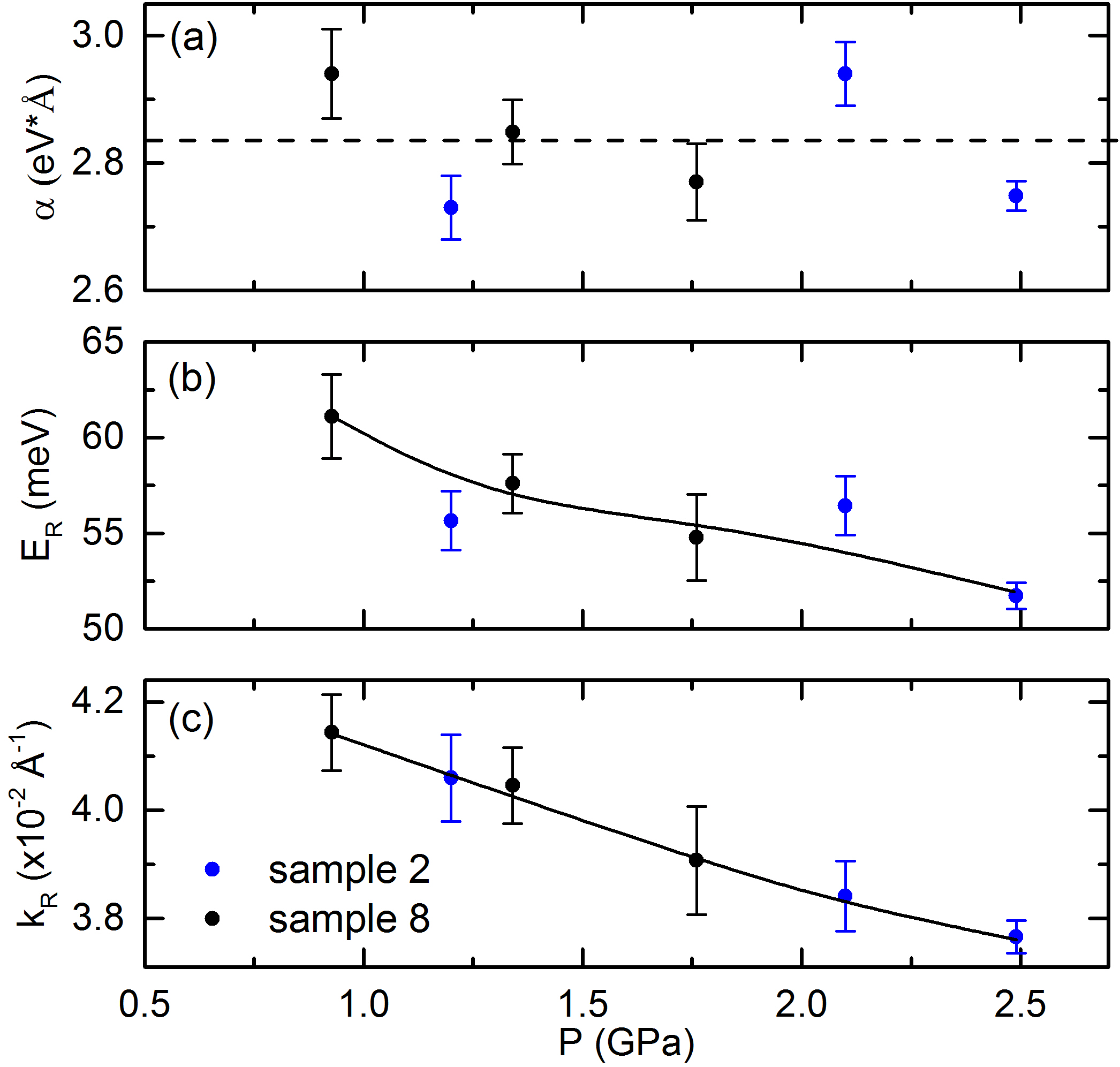}
    \caption{Rashba parameters as functions of pressure for samples 2 and 8. (a) The Rashba coupling, $\alpha$, stays roughly constant, (b) the Rashba energy and (c) the Rashba momentum both decrease with pressure. The curves in (b) and (c) are guides to the eye.}
 \label{fig4}
\end{figure}

With the aid of a simple, two-band model Hamiltonian discussed in Ref.~\cite{vangennep_2014_1}, we use the cyclotron masses and frequencies to extract several parameters which describe the spin-split conduction bands of BiTeCl. Figure~\ref{fig4} shows the calculated values of the Rashba coupling, $\alpha$, the Rashba energy, $E_R$, and the Rashba momentum, $k_R$, as a function of pressure. We observe that $\alpha$ stays roughly constant, while both $E_R$ and $k_R$ decrease with an increase in applied pressure. 

\begin{figure}
  \raggedleft
    \includegraphics[width=\columnwidth]{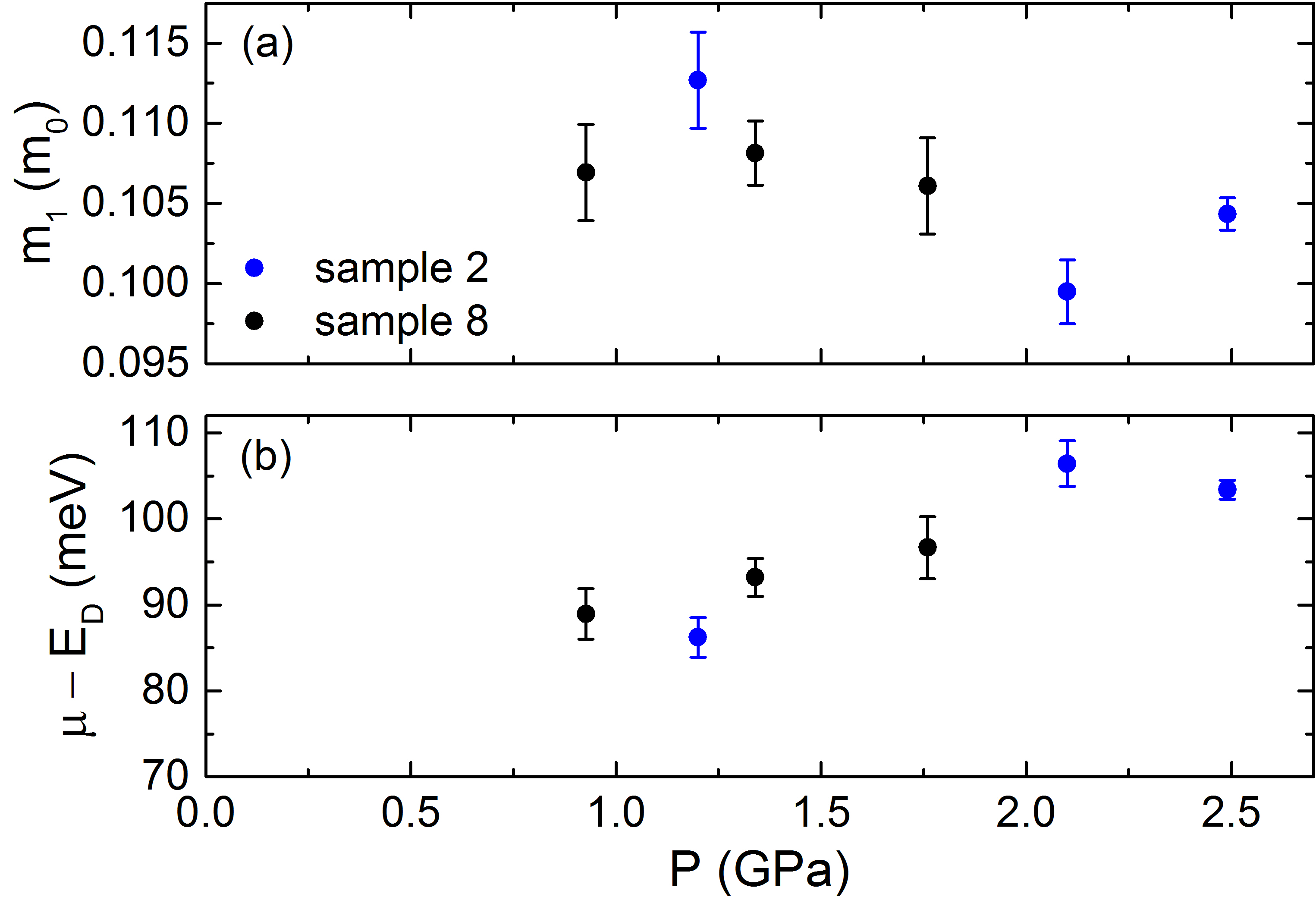}
    \caption{Calculated band parameters for samples 2 and 8 at 2 K as functions of pressure. (a) Band mass and (b) chemical potential referenced to the Dirac point.}
 \label{fig5}
\end{figure}

The model also allows us to calculate the band mass, $m_1$, as well as the position of the chemical potential, $\mu$, relative to the Dirac point, $E_D$. Fig.~\ref{fig5} shows the $m_1$ and $\mu$ as a function of applied pressure. We note that $m_1$ has a slight trend downward with pressure, but does not change significantly. In contrast, $\mu$ rises at a rate of $\sim12~ \mathrm{meV/GPa}$, relative to the Dirac point.

The results from this study on BiTeCl show similar trends in the pressure dependences of the band parameters when compared to a previous study on BiTeI ~\cite{vangennep_2014_1}. Most notably, we find that $\alpha$ stays roughly constant for both materials up to $\sim 2.5\,\mathrm{GPa}$. We also find that the  rate of change of the chemical potential $\mu$ with pressure is similar in both materials; $\mu$ rises at a rate of $20(1),\mathrm{meV/GPa}$ for BiTeI, and $12(3)\, \mathrm{meV/GPa}$ for BiTeCl. In both materials, we find that the splitting of the bands, measured by $k_R$, decreases with pressure at nearly the same rate: $-0.0025(3)~{\mbox{\AA}^{-1}/\mathrm{GPa}}$ for BiTeI, and $-0.0028(3)~{\mbox{\AA}^{-1}/\mathrm{GPa}}$ for BiTeCl. The band masses for both materials are roughly constant, with perhaps a slight reduction with pressure.

A recent optical study of BiTeCl under pressure shows a transition with an energy of $\sim 200\,\mathrm{meV}$, which is attributed to excitations between the Rashba-split conduction sub-bands~\cite{Crassee2017}. Using the parameters extracted from our measurements and the parabolic band model~\cite{vangennep_2014_1}, we estimate this interband transition energy to be $\sim 600\,\mathrm{meV}$. The discrepancy between the two values is likely due to a non-parabolicity of the bands well above the chemical potential that is not captured by the model that we have used.

\section{Conclusions}
In summary, we have performed measurements of Shubnikov-de Haas oscillations in BiTeCl at several pressures up to $\sim 2.5\,\mathrm{GPa}$.  Using a simple parabolic model of the Rashba split bands~\cite{vangennep_2014_1} together with the effective masses and oscillation frequencies determined from our measurements, we are able to estimate how the conduction bands evolve with pressure.  Overall, the pressure dependencies of the band parameters in BiTeCl are very similar to those observed in a previous study of BiTeI~\cite{vangennep_2014_1}. 

Despite the slightly different ambient pressure crystal structures adopted by BiTeI and BiTeCl, it is not unreasonable to view the replacement of the larger I ion with the smaller Cl ion as resulting in chemical pressure.  This is consistent with BiTeI exhibiting band inversion only under applied pressure, while in BiTeCl the bands are inverted at ambient pressure.  Accordingly, the similar results found in this study on BiTeCl and our previous study on BiTeI~\cite{vangennep_2014_1} suggest that the band inversion and trivial to non-trivial topological transition has little influence on the pressure evolution of the bulk conduction electrons. Finally, we find no clear sign of a bulk bandgap closure, suggesting that the topological phase of BiTeCl is robust under pressures up to $\sim 2.5\, \mathrm{GPa}$. 

\section*{Acknowledgments}
DV, DEJ, and JJH acknowledge support from the National High Magnetic Field Laboratory's User Collaborative Grants Program (UCGP) as well as support from the National Science Foundation (NSF) from grant NSF DMR-1453752.  DG acknowledges support from the Department of Energy (DOE) from grant DOE NNSA DE-NA0001979. The National High Magnetic Field Laboratory is supported by the NSF via Cooperative agreement  No.\ DMR-1157490, the State of Florida, and the U.S. Department of Energy. 

\section*{References}
\providecommand{\newblock}{}

\end{document}